\documentclass[conference]{IEEEtran}

\usepackage[algo2e,ruled,vlined]{algorithm2e}
\SetCommentSty{scriptsize}
\SetKwComment{tcc}{\{}{\}}
\SetKwIF{If}{ElseIf}{Else}{if}{}{else if}{else}{endif}
\SetKwFor{While}{while}{}{endw}
\SetKwFor{ForEach}{for each}{}{endfch}
\SetKwBlock{Par}{parallel region}{}

\SetKwRepeat{Do}{do}{while}

\usepackage{subcaption}
\usepackage{graphicx}
\usepackage{url}

\title{Persistent and Partitioned MPI for Stencil Communication}

\author{
\IEEEauthorblockN{Gerald Collom}
\IEEEauthorblockA{\textit{Department of Computer Science} \\
\textit{University of New Mexico}\\
Albuquerque, USA \\
geraldc@unm.edu}
\and
\IEEEauthorblockN{Jason Burmark}
\IEEEauthorblockA{\textit{Lawrence Livermore National Lab} \\
Livermore, USA \\
burmark1@llnl.gov}
\and
\IEEEauthorblockN{Olga Pearce}
\IEEEauthorblockA{\textit{Lawrence Livermore National Lab} \\
Livermore, USA \\
pearce8@llnl.gov}
\and
\IEEEauthorblockN{Amanda Bienz}
\IEEEauthorblockA{\textit{Department of Computer Science} \\
\textit{University of New Mexico}\\
Albuquerque, USA \\
bienz@unm.edu}
}

\begin{document}

\maketitle

\begin{abstract}
    Many parallel applications rely on iterative stencil operations, whose performance are dominated by communication costs at large scales.
    Several MPI optimizations, such as persistent and partitioned communication, reduce overheads and improve communication efficiency through amortized setup costs and reduced synchronization of threaded sends.  This paper presents the performance of stencil communication in the Comb benchmarking suite when using non-blocking, persistent, and partitioned communication routines.  The impact of each optimization is analyzed at various scales.  Further, the paper presents an analysis of the impact of process count, thread count, and message size on partitioned communication routines.  Measured timings show that persistent MPI communication can provide a speedup of up to $37\%$ over the baseline MPI communication, and partitioned MPI communication can provide a speedup of up to $68\%$.
\end{abstract}

\begin{IEEEkeywords}
 MPI, stencil exchanges, communication, persistent communication, partitioned communication
\end{IEEEkeywords}

\section{Introduction}~\label{sec:intro}
Stencil computations dominate a wide range of parallel applications, including scientific simulations based on solving partial differential equations~\cite{abhyankar2018petsctsmodernscalableodedae} and numerical solvers such as geometric multigrid~\cite{7097787}. 
They are established as one of the most prevalent patterns in high performance computing~\cite{10.1145/1562764.1562783}.
Stencil operations consist of solving a linear method across a structured mesh.  
At each iteration, all mesh cells are updated, with the new value at each cell dependent on that of all neighboring cells.  
In parallel, the mesh is decomposed into subdomains for each participating process, with each process holding boundary cell data from processes with neighboring mesh regions (ghost cells) in order to compute its own boundary cell updates. 
A parallel stencil operation consists of updating values locally before exchanging updated ghost regions during each iteration.  
As process counts increase, these boundary exchanges dominate performance costs.  
As a result, efficient boundary exchanges are crucial for performant and scalable stencil codes.

There are several common implementations for boundary exchanges, including hand-rolled point-to-point communication and neighborhood collectives with Cartesian process grids. 
Regardless of approach, potential optimizations for boundary exchanges have been introduced with each major release of the MPI standard.
For example, persistent communication introduced in MPI 3 has shown to improve communication costs~\cite{hatanaka13, georg2017pmr, jammer22} by incurring setup and overhead costs once while allowing cheaper iterations of communication afterward.  
Furthermore, MPI may choose to optimize persistent exchanges through additional overheads in the initialization, such as tag matching.
MPI 4 provides an additional optimization opportunity through partitioned communication, allowing threads to concurrently work on a persistent message.  This paper presents a novel study of the trade-offs between standard non-blocking, persistent, and partitioned communication throughout iterative boundary exchanges.

Comb, a stencil benchmarking tool from Lawrence Livermore National Laboratory (LLNL) analyzes the performance trade-offs between various stenciled boundary exchanges.  This work adds both persistent and partitioned MPI optimizations into this widely-used benchmark to analyze their impact.  The optimizations are made publicly available within Comb for application developers to use when analyzing comparable communication strategies.

The contributions of this paper include the following.
\begin{enumerate}
    \item Optimize stencil exchanges with persistent communication
    \item Further optimize stencil exchanges with partitioned communication
    \item Demonstrate up to $68\%$ speedup over existing MPI methods in Comb when scaled across thousands of processes.
    \item A performance study of persistent and partitioned communication for stencil exchanges across varying process, CPU core and thread counts.
    \item Expand the stencil communication benchmark Comb to allow application programmers to test persistent and partitioned communication strategies.
\end{enumerate}

The following sections are organized as follows. Section~\ref{sec:bkgrnd} provides background on stencil codes and the persistent and partitioned communication optimizations.
Related works are discussed in Section~\ref{sec:related}.
In Section~\ref{sec:impl}, the implementation of the two communication optimizations are detailed.
Results of performance analysis of persistent and partitioned MPI in the stencil communication benchmark Comb are provided in Section~\ref{sec:results}.
Finally, conclusions and future directions are presented in Section~\ref{sec:conc}.

\section{Background}~\label{sec:bkgrnd}
Stencil codes split a structured mesh across a Cartesian process grid, requiring processes to iteratively perform local computation followed by boundary exchanges.  
Each process uses initial boundary data for each neighboring process to update all local mesh cells.  
After local updates are complete, local boundary cell values are packed into a contiguous buffer and communicated to neighboring processes.
Boundary cell values from neighboring process are also received and unpacked from a contiguous buffer into the local mesh.
After all updates of ghost cells are complete, successive iterations can occur.
This exchange of data, also known as a halo exchange, is further detailed in Section~\ref{sec:halo}.

\subsection{Halo Exchanges}~\label{sec:halo}
The pattern of boundary data to be exchanged at each iteration is dependent on the structure of the problem at hand.
\begin{figure}
    \centering
    \includegraphics[width=\linewidth]{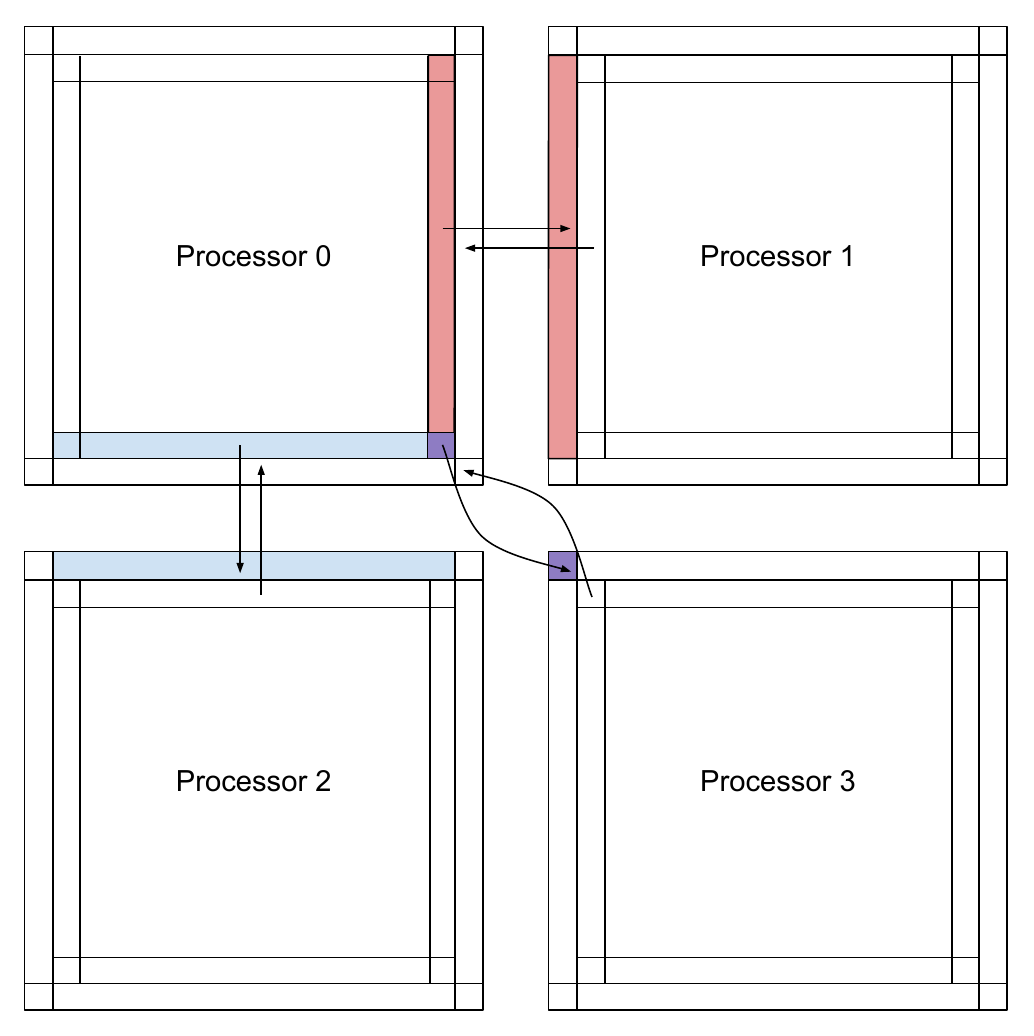}
    \caption{Shown are example exchanges for a single process in a 2D regular halo exchange. Process 0 sends values for every cell along its right edge (red and purple) to Process 1 to fill its left edge of ghost cells (red). All cells along the bottom edge (blue and purple) are sent to Process 2 to fill its top edge of ghost cells (blue). Finally, the purple corner is sent to Process 3 to fill its top left ghost cell. In return, Processes 1, 2 and 3 send their boundary cell values (white) to Process 0 to update its ghost cells.}
    \label{fig:halo}
\end{figure}
For example, a 2D stencil requires exchanging edges and corners with up to 8 neighbors, as exemplified in Figure~\ref{fig:halo}.  
In 3D, this is often extended to a 27-point stencil, in which faces, edges, and corners are communicated to 26 neighboring processes.  

While some data is contiguous, such as boundaries on the top and bottom of each process in a row-wise data layout, other exchanges require communication of non-contiguous data, such as the left and right edges.  
Multiple methods for exchanging non-contiguous data have been explored, including using MPI Datatypes~\cite{hashmi2020falcon, pearson2021tempi}.  
However, on CPU-only supercomputers, packing costs are typically outweighed by communication at scale.  
As a result, this work is focused on optimizing the exchange of data, and simply packs each message into a contiguous buffer outside MPI before communication.
OpenMP threads are used to perform packing, allowing the benefit of utilizing all CPU cores.

\begin{algorithm2e}
    \DontPrintSemicolon%
    \KwIn{$\texttt{msgs\_info, requests}$}
    \tcc*{message information (process pairs, data sources), requests storage array}
    \BlankLine%
    \Par(\tcc*[f]{OpenMP Region}){
        \tcp{Copy data from mesh to contiguous message buffer}
        \texttt{pack}($\texttt{msgs\_info}$)\;
    }
    \BlankLine%
    
    \tcp{Begin communication}
    \For{$i \gets 0$ \KwTo $n_{send}$}{ 
       MPI\_Isend(msgs\_info[$i$], requests[i])\;
    }
    \For{$i \gets 0$ \KwTo $n_{recv}$}{ 
       MPI\_Irecv(msgs\_info[$i + n_{send}$], requests[$i + n_{send}$])\;
    }
    
    \BlankLine%
    \tcp{Wait on messages}
    \texttt{MPI\_Waitall(requests)}
    
    \BlankLine%
    \Par {
        \tcp{Copy values from message buffer into mesh}
        \texttt{unpack}($\texttt{msgs\_info, msgs\_info}$)\;
    }
    
    \caption{\texttt{Exchange}}\label{alg:exchange}
\end{algorithm2e}
A standard boundary exchange is detailed in Algorithm~\ref{alg:exchange}.  
All messages are first packed into a contiguous buffer using OpenMP threads, before being exchanged with neighboring processes.  
Finally, data is unpacked into the local domain, to be used during successive steps of computation.

\subsection{Persistent MPI and Partitioned MPI}

Persistent communication allows communication patterns to be defined and initialized once and repeatedly executed to avoid repeated overhead costs. 
This optimization is well suited for applications like stencil codes with iterative communication.
A persistent communication pattern consists of calling a method such as \texttt{MPI\_Send\_init} once to initialize communication, providing all arguments that are typically provided to \texttt{MPI\_Isend}. 
However, rather than beginning communication, \texttt{MPI\_Send\_init} returns a handle for a persistent request.
Each iteration of communication is then performed by passing the persistent request to persistent exchange routines, such as \texttt{MPI\_Start} and \texttt{MPI\_Wait}.
As a result, initialization costs are only incurred once and then amortized over all subsequent exchanges.

Partitioned communication builds upon persistent communication, partitioning each message across multiple threads and allowing for portions of a message to be independently sent and received without synchronizing on the entirety of the message.
This optimization allows threads to work concurrently and for communication of parts of a large message to begin as soon as they are ready, without waiting for the entire message data to be ready.
This early communication can reduce network contention by utilizing the network early rather than sending all data at once.
Further, partitioned communication allows for early work, in which successive computation (or unpacking) can begin as soon as any portion of a message arrives.

Partitioned communication can work well for large messages, but does incur some overheads.
The flag \texttt{MPI\_THREAD\_MULTIPLE} is required as multiple threads all perform communication at the same time, which can cause slowdowns that vary greatly among versions of MPI~\cite{zambre22}.
Additionally, for a given message, all partitions must be equal in size and there must be an equal number of partitions on the sending and receiving sides.
While padding can be added to the last message for those with sizes that do not evenly divide partition count, this padding can be complex.  

Similar to persistent communication, partitioned sends first require an initialization call, \texttt{MPI\_Psend\_init}, requiring standard communication arguments along with the number and size of partitions for subsequent iterations.
Each iteration, data is exchanged through a single instance of \texttt{MPI\_Pstart} per message, followed by invoking \texttt{MPI\_Pready} for each partition to mark the portion of data is ready to be sent.
Finally, \texttt{MPI\_Wait} is called to await the arrival of all partitions.
If attempting to perform early work, \texttt{MPI\_Parrived} can be used to test the arrival of an individual partition, which, if arrived, is ready for use in computation.

\section{Related Works}~\label{sec:related}

Persistent MPI communication was introduced in the MPI 1.1 standard.
Since then, work has been done to implement and optimize persistent communication~\cite{hatanaka13, jammer22}.
As an older optimization, adoption is not uncommon in practice, although far from ubiquitous.
Persistent communication has also been used within the context of neighborhood collectives.~\cite{morgan17}

Several works have tackled the design of partitioned communication within MPI~\cite{partitioned, partitionedmpi4, bangalore20, worley21, marts23}.
Additional works have explored methods to accurately model and benchmark partitioned communication, and analyzed the potential benefits of this optimization~\cite{partitionedmpi4, hassan22, schonbein23, gillis23}.
Partitioned MPI communication has also been evaluated in comparison to other approaches to MPI+Threads~\cite{zambre22}.
The combination of partitioned and collective communication is another direction that has previously been explored~\cite{holmes21}.

Separately, prior work looked into the optimization of stencil code communication without the use of persistent or partitioned MPI~\cite{pearson20, pekkila22}.
Specifically, optimization techniques such as node-awareness have been successfully applied to stencil communication~\cite{pearson20}.
MPI Datatypes have also seen use in stencil computations~\cite{hashmi2020falcon, pearson2021tempi}.
The stencil communication benchmark Comb can also test on heterogeneous architectures and prior work has explored optimizations in this case~\cite{elis2024non}, although the work in this paper focuses on CPU only systems.

\section{Implementations}~\label{sec:impl}

Standard stencil codes use non-blocking sends and receives at each iteration.
Both persistent and partitioned communication optimizations are suitable for this structured, iterative exchange.
Because the stencil computation and communication is iterative, the use of persistent communication provides a natural optimization to incur an overhead cost once at initialization and utilize it in every following iteration of communication.
In the case where multiple threads are used to efficiently pack data into contiguous buffers, partitioned communication is a natural fit, as these same threads can be used to parallelize communication with partitions.
Furthermore, exchanges of entire faces of a 3D domain often result in very large messages, which can be optimized when split across many available CPU cores and communicated asynchronously.

\subsection{Persistent MPI}

\begin{algorithm2e}
    \DontPrintSemicolon%
    \KwIn{$\texttt{msgs\_info, requests}$}
    \tcc*{message information (process pairs, data sources), requests storage array}
    \BlankLine%

    \tcp{Initialize messages}
    \For{$i \gets 0$ \KwTo $n_{send}$}{ 
       \texttt{MPI\_Send\_init}(\texttt{msgs\_info[$i$], requests[$i$]})\;
    }
    \For{$i \gets 0$ \KwTo $n_{recv}$}{  
       \texttt{MPI\_Recv\_init}(\texttt{msgs\_info[$i + n_{send}$], requests[$i + n_{send}$]})\;
    }
    
    \caption{\texttt{Persistent Init}}\label{alg:pers_init}
\end{algorithm2e}

\begin{algorithm2e}
    \DontPrintSemicolon%
    \KwIn{$\texttt{msgs\_info, requests}$}
    \tcc*{message information (process pairs, data sources), requests storage array}
    \BlankLine%
    \Par(\tcc*[f]{OpenMP Region}){
        \tcp{Copy data from mesh to contiguous message buffer}
        \texttt{pack}($\texttt{msgs\_info}$)\;
    }
    \BlankLine%
    
    \tcp{Begin communication}
    \texttt{MPI\_Startall(requests)}
    
    \BlankLine%
    \tcp{Wait on messages}
    \texttt{MPI\_Waitall(requests)}
    
    \BlankLine%
    \Par{
        \tcp{Copy values from message buffer into mesh}
        \texttt{unpack}($\texttt{msgs\_info}$)\;
    }
    
    \caption{\texttt{Persistent Exchange}}\label{alg:pers_exchange}
\end{algorithm2e}

\begin{algorithm2e}
    \DontPrintSemicolon%
    \KwIn{$\texttt{requests}$\tcc*{requests storage array}}
    \BlankLine%
    \For{$i \gets 0$ \KwTo $n_{recv} + n_{send}$}{
        \texttt{MPI\_Request\_free(requests[$i$])}\;
    }
    \caption{\texttt{Persistent Destroy}}\label{alg:pers_destroy}
\end{algorithm2e} 

Persistent stencil communication is split into three separate methods: initialization, iterative exchange, and destruction.
First, each process initializes the persistent exchange, as described in Algorithm~\ref{alg:pers_init}.
This consists of initializing each persistent send and receive.
Then, during each iteration of the persistent boundary exchange described in Algorithm~\ref{alg:pers_exchange}, all processes pack data in an equivalent fashion to the standard approach in Algorithm~\ref{alg:exchange}.
However, all non-blocking communication is now replaced with \texttt{MPI\_Startall} and \texttt{MPI\_Waitall}.
Finally, when all iterations are complete, the persistent request handles must be destroyed, as shown in Algorithm~\ref{alg:pers_destroy}.

\subsection{Partitioned MPI}
\begin{algorithm2e}
    \DontPrintSemicolon%
    \KwIn{$\texttt{n\_parts, msgs\_info, requests}$}
    \tcc*{number of partitions, message information (process pairs, data sources), requests storage array}
    \BlankLine%

    \tcp{Initialize messages}
    \For{$i \gets 0$ \KwTo $n_{send}$}{ 
       \texttt{MPI\_Psend\_init}(\texttt{n\_parts, msgs\_info[$i$], requests[$i$]})\;
    }
    \For{$i \gets 0$ \KwTo $n_{recv}$}{  
       \texttt{MPI\_Precv\_init}(\texttt{n\_parts, msgs\_info[$i + n_{send}$], requests[$i + n_{send}$]})\;
    }
    
    \caption{\texttt{Partitioned Init}}\label{alg:part_init}
\end{algorithm2e}

\begin{algorithm2e}
    \DontPrintSemicolon%
    \KwIn{$\texttt{msgs\_info, requests}$}
    \tcc*{message information (process pairs, data sources), requests storage array}
    \BlankLine%
    \tcp{Begin communication}
    \texttt{MPI\_Startall(requests)}
    \BlankLine%

    \Par(\tcc*[f]{OpenMP Region}){
        \tcp{Copy data from mesh to contiguous message buffer}
        \texttt{pack}($\texttt{msgs\_info}$)\;
        \tcp{Mark partition of current thread as ready}
        \texttt{MPI\_Pready}(\texttt{partition})\;
    }
    \BlankLine%
    
    \tcp{Wait on messages}
    \texttt{MPI\_Waitall(requests)}

    \BlankLine%
    \Par{
        \tcp{Copy values from message buffer into mesh}
        \texttt{unpack}($\texttt{msgs\_info}$)\;
    }
    
    \caption{\texttt{Partitioned Exchange}}\label{alg:part_exchange}
\end{algorithm2e}

\begin{algorithm2e}
    \DontPrintSemicolon%
    \KwIn{$\texttt{requests}$\tcc*{requests storage array}}
    \BlankLine%
    \For{$i \gets 0$ \KwTo $n_{recv} + n_{send}$}{
        \texttt{MPI\_Prequest\_free(requests[$i$])}\;
    }
    \caption{\texttt{Partitioned Destroy}}\label{alg:part_destroy}
\end{algorithm2e} 

Partitioned communication further optimizes stencil operations beyond the persistent optimizations, allowing for threads to each communicate a portion of the data.
This optimization consists of first initializing the partitioned exchange, as described in Algorithm~\ref{alg:part_init}, during which partitioned MPI initialization calls are instantiated for each message.
The arguments provided are the same in Algorithm~\ref{alg:pers_init}, except an additional argument for the number of partitions of the message.
Then, every iteration of the partitioned stencil code performs a boundary exchange among all threads, as shown in Algorithm~\ref{alg:part_exchange}.
In this method, the packing thread calls \texttt{MPI\_Pready} immediately after packing its portion of the data, marking the specified message partition as ready to send.  
For simplicity, all process then wait for all messages to complete before unpacking.  
However, this approach could be further optimized through early work, in which each thread could unpack partitions of a message on arrival.
After all iterations of halo exchanges have completed, the partitioned request handle is destroyed, as shown in Algorithm~\ref{alg:part_destroy}.

\section{Results}~\label{sec:results}

All optimizations described in Section~\ref{sec:impl} have been added throughout the Comb halo exchange benchmarking suite from LLNL. 
The partitioned communication implementation we used was the MPIX Partitioned Communication Library (MPIPCL)\footnote{https://github.com/mpi-advance/MPIPCL} and included it directly within Comb.
The presented timings are acquired from Comb simulations with three mesh variables, a single cell wide exchange boundary and a periodic problem mesh in all directions so that no special case edge regions occur in the decomposition.
Additionally, the following Comb options were disabled: \texttt{per\_message\_pack\_fusing} and \texttt{message\_group\_pack\_fusing}.
All presented timings account for asynchrony of processes and timer inaccuracies through multiple methods.  
Before any timing is initialized, a barrier occurs to synchronize all processes.  
Further, timer precision is accounted for, and all measurements are timed over $1000$ exchanges, with the average per-exchange cost extracted.
Finally, the impact of nearby jobs is lessened as each test is performed three separate times on the system and the average time of the three runs is used.

Standard, persistent, and partitioned exchanges are analyzed on the Quartz supercomputer, an Intel SMP architecture at LLNL, using the default system MPI Mvapich2 version 2.3.7.  
While Quartz nodes each contain $36$ cores, all performance measurements in this section use $32$ processes per node to keep all process counts powers of $2$.  
Further, all tested thread counts are also powers of $2$, and all stencil sizes were chosen such that halo exchange sizes are also powers of $2$.  
As a result, partitioned communication is performed in all cases without the need for padding, but it should be noted that to use partitioned communication as currently defined by the MPI standard, padding would need to be added to account for uneven partition sizes.  
Finally, our tests used 2 OpenMP threads per core to take advantage of hyperthreading.

\begin{figure}[h]
    \centering
    \includegraphics[width=\linewidth]{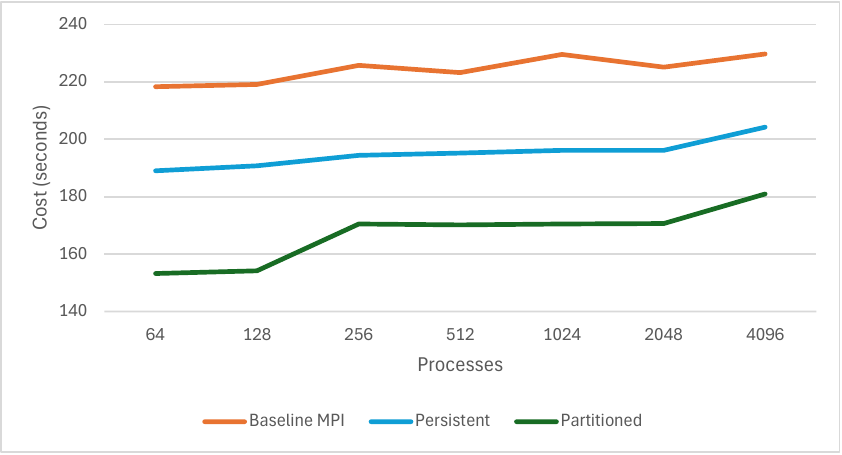}
    \caption{Weak scaling of the tested communication strategies for message sizes of $524,288$ doubles.}
    \label{fig:weak}
\end{figure}
In Figure~\ref{fig:weak}, we present a weak scaling study from $64$ to $4096$ processes, with messages sizes of $524,288$ double-precision floats (doubles) at each scale.  
All cases in this study are run with $32$ ranks per node, $32$ active cores per node, and $2$ threads per core.  
At the largest scale of $4096$ processes, persistent communication obtains a $12.5\%$ speedup over Comb's baseline MPI implementation, while partitioned communication further improves performance by another $14.5\%$ to a total of $27\%$ over the baseline.  
Note, performances for all three methods trend upwards at higher scales and appear to slightly approach each other, suggesting additional communication or contention costs occur at larger scales.

\begin{figure}[h]
    \centering
    \includegraphics[width=\linewidth]{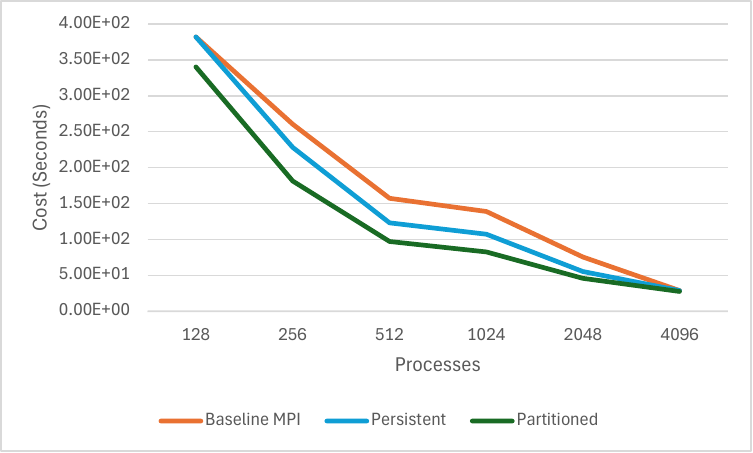}
    \caption{Strong scaling study of the communication strategies for a $2048^3$ cell mesh. Message sizes start at $262,144$ doubles.}
    \label{fig:strong}
\end{figure}

In Figure~\ref{fig:strong}, we present a strong scaling study of the communication costs of the different policies from $128$ to $4096$ processes, using equivalent per-node process, core, and thread counts to the weak scaling study.  
This test simulates a $2048^3$ cell problem, which results in message sizes of $262,144$ doubles at the smallest scale, decreasing as the number of processes grows.  
Persistent communication achieves speedup of $37\%$ over the baseline at $2048$ processes, but performs equivalently to standard approaches at both the smallest and largest process counts.  
Partitioned communication outperforms the other methods, achieving speedups over the baseline of $12\%$, $68\%$, and $4.4\%$ at 128, 1024, and 4096 processes respectively.  
Note, partitioned communication provides less of an advantage for smaller message sizes.  
This is a contributing factor to the diminishing speedup from the use of partitioned communication at higher process scales for the same problem size.

\begin{figure}[h]
    \centering
    \includegraphics[width=\linewidth]{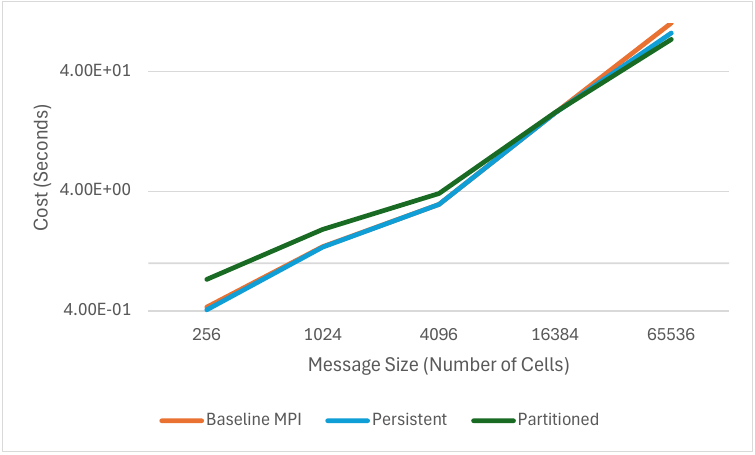}
    \caption{A comparison of the communication strategies on $4096$ processes as message size is increased from $768$ to $196608$ doubles.}
    \label{fig:msg_size}
\end{figure}

Figure~\ref{fig:msg_size} displays the cost of halo exchanges with each of the three policies on $4096$ processes, with varying halo exchange sizes.  
The range of message sizes varies from $256$ to $65536$ mesh cells consisting of 3 doubles each.  
At the smallest message scale tested, persistent communication performed similarly to the baseline, while partitioned communication performs significantly worse, with the baseline performing $73\%$ faster.
As message sizes increase, however, persistent and partitioned communication performance eventually overtake for a speedup over the baseline of $21\%$ for persistent and $37\%$ for partitioned at the largest tested message size.

\begin{figure}[h]
    \centering
    \includegraphics[width=\linewidth]{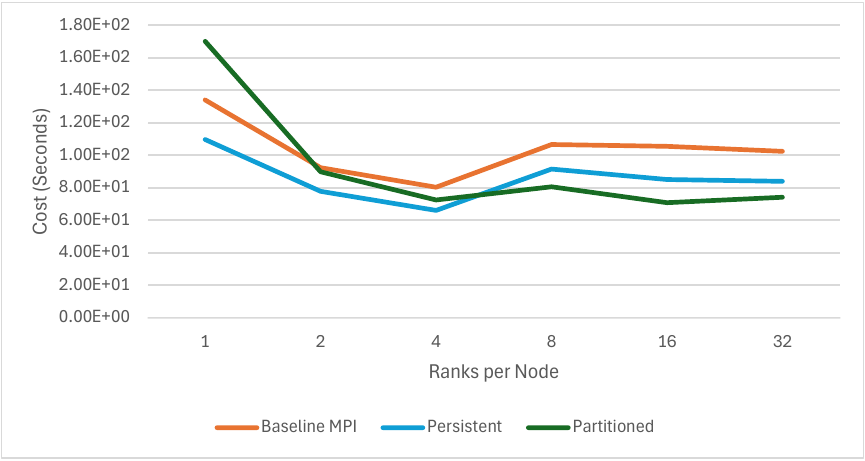}
    \caption{A study of the effect of the number of MPI ranks used per node on the performance of the communication strategies for a $2048 \times 4096 \times 4096$ mesh.}
    \label{fig:cores}
\end{figure}
Figure~\ref{fig:cores} shows the cost of halo exchanges across $64$ nodes, with varying numbers of MPI processes per node.  
All cases consist of $32$ active cores per node and $64$ OpenMP threads per node.  
As the number of MPI ranks per node increases, OpenMP threads per rank decrease accordingly.  
As a result, when using fewer MPI processes per node, thread launch overheads increase.  
Furthermore, in the standard and persistent cases, fewer cores are utilized during MPI communication.  
All test contain a constant problem size of $2048 \times 4096 \times 4096$ cells.

This study shows that persistent communication outperforms the baseline for every tested number of ranks per node, with a speedup of around $20\%$ in each case.  
Additionally, at one rank per node, partitioned communication performs significantly worse than other methods, including the baseline.  
At 2 ranks per node, however, partitioned communication performance slightly overtakes the baseline and at 8 ranks per node, it overtakes persistent communication performance as well.  
As the ranks per node increase, the threads per rank decrease.  
Therefore, partitioned communication only achieves speedups when each rank has a limited number of threads, indicating that there is a limit to the number of partitions into which a message should be split.  
Note for the single rank per node case, the threads were likely split across sockets, potentially amplifying overheads.

\section{Conclusions and Future Directions}~\label{sec:conc}

This paper demonstrated how persistent and partitioned MPI communication optimizations can be used to improve halo exchanges.  
Persistent communication optimizes iterative stencil exchanges, providing significant speedups over standard non-blocking communication.  
Partitioned communication provides additional speedups when fewer threads are used per process, as long as data exchanges are large.

Additional partitioned communication optimizations can be further explored in the future, namely performing early work, unpacking data immediately upon arrival through the use of the \texttt{MPI\_Parrived} routine.  
Furthermore, the use of partitioned communication can be explored on heterogeneous architectures.  
Depending on future availability of stream and kernel triggered MPI routines, it is possible partitioned communication could be used to optimize GPUDirect communication.  
Partitioned communication could also be explored in the context of copy-to-CPU communication routines on heterogeneous architectures, as modeling results have shown use cases GPUDirect is outperformed when copying to the CPU and using all available CPU cores~\cite{bienz2021modeling}.  
Partitioned communication has the potential to further improve copy-to-CPU methods by partitioning large messages across all available CPU cores.  

This paper presents an initial analysis of the impact of partitioned communication on stencil exchanges, with results guiding use cases for the optimization.  
However, each application programmer is currently required to implement these optimizations by hand to achieve the performance benefits shown in this paper.  
The work presented throughout this paper could, in the future, be made accessible through the use of persistent neighborhood collectives, allowing the underlying system MPI to choose the optimal approach for a given iterative halo exchange.

\section*{Acknowledgment}
This work was performed with partial support from the National Science
Foundation under Grant No. CCF-2151022 and the U.S. Department of Energy's National Nuclear Security Administration (NNSA) under the Predictive Science Academic Alliance Program (PSAAP-III), Award DE-NA0003966.

Any opinions, findings, and conclusions or recommendations expressed in this material are those of the authors and do not necessarily reflect the views of the National Science Foundation and the U.S. Department of Energy's National Nuclear Security Administration.

\bibliographystyle{siamplain}
\bibliography{references}
\end{document}